%% file: main.tex
\documentclass[11pt,a4paper]{article}

\pdfoutput=1 

\usepackage{jinstpub} 
\usepackage[]{graphicx}
\usepackage{subfig}
\usepackage{slashed}
\usepackage{todonotes}
\usepackage{lineno}

\input{mycommands.tex}

\let\oldequation\equation
\let\oldendequation\endequation

\renewenvironment{equation}
  {\linenomathNonumbers\oldequation}
  {\oldendequation\endlinenomath}

\makeatletter
\renewcommand{\todo}[2][]{
    \@todo[caption={#2}, #1]{\tiny \setstretch{1.}#2}%
}
\makeatother

\title{\boldmath Rebalance and Smear for Multi-jet Background Estimation}
\author[1]{Samuel Bein}
\author[2]{Christian Sander}
\affiliation[1]{\textit{\small Universit\"at Hamburg, Luruper Chaussee 149, 22761 Hamburg, Germany}}
\affiliation[2]{\textit{\small Deutsches Elektronen-Synchrotron DESY, Notkestr. 85, 22607 Hamburg, Germany}}
\emailAdd{samuel.bein@cern.ch}
\date{\today}

\abstract{
For many particle collider searches for physics beyond the Standard Model in final states with jets and missing transverse momentum, events from QCD multi-jet processes are an important and challenging background contribution. The CMS and ATLAS experiments have previously developed data-driven methods designed to succeed where Monte Carlo methods suffer large theoretical and experimental uncertainties. One such method is \textit{Rebalance and Smear} (R\&S), which predicts QCD backgrounds by applying a series of folding and unfolding techniques to data control regions. A top-to-bottom description of the R\&S method is presented, along with a discussion of its applicability and limitations. A software application is provided that performs the R\&S method using public, non-proprietary tools, interfacing with data sets produced by \texttt{Delphes3}. In general, the method is suitable for predicting multi-jet backgrounds in searches for stable non-detectable particles, such as dark matter candidates. A case study is carried out in simulated events of proton-proton collisions at $\sqrt{s}=14$ TeV in the context of a potential search for Higgsino dark matter produced in the decay products of supersymmetric quark partners.
Sources of potential bias are explored and prescriptions for evaluating systematic uncertainties are suggested.
}

\begin{document}

\maketitle
\nolinenumbers


\input{intro.tex}
\input{method.tex}
\input{example.tex}
\input{uncertainties.tex}
\input{summary.tex}

\bibliography{master}

\appendix

\input{appendix_code.tex}

\end{document}

%% file: mycommands.tex
\newcommand{\TeV}{\ensuremath{\text{TeV}}}
\newcommand{\GeV}{\ensuremath{\text{GeV}}}

\newcommand{\met}{\ensuremath{\slashed{p}_{\text{T}}}}
\newcommand{\metvec}{\ensuremath{\slashed{\boldsymbol{p}}_{\text{T}}}}
\newcommand{\methard}{\ensuremath{\slashed{p}_{\text{T}}^{\text{hard}}}}
\newcommand{\metvechard}{\ensuremath{\boldsymbol{\slashed{p}}_{\text{T}}^{\text{hard}}}}

\newcommand{\HT}{\ensuremath{H_{\text{T}}}}

\newcommand{\pt}{\ensuremath{p_{\text{T}}}}

%% file: intro.tex
\section{Motivation}
\label{sec:intro}

The possibility of observing new particles beyond the Standard Model (SM) at the energy frontier is the subject of much investigation, particularly at the Large Hadron Collider (LHC)~\cite{Evans:2008zzb}.
A leading factor motivating this effort is the observation of dark matter (DM) as the dominant contribution to the mass of galaxies and the cosmic medium~\cite{Zwicky:1937zza,Rubin:1970zza,Clowe:2006eq}.
A rich programme of DM searches is performed at the LHC, in particular by the ATLAS~\cite{Aad:2008zzm} and CMS~\cite{Chatrchyan:2008aa} experiments, and large progress has been made in excluding parameter regions of various models identified as consistent with the dark matter hypothesis~\cite{Boveia:2018yeb}, which serve as complementary constraints to direct detection experiments~\cite{Aprile:2018dbl,Cui:2017nnn,Akerib:2016vxi,Agnes:2018ves}.

The sensitivity of the LHC experiments to such a production of DM is maximal when considering scenarios in which new, strongly-interacting particles are produced and each decay into one or more jets and a DM particle.
The most prominent example for an extension of the SM that allows for this topology is supersymmetry (SUSY)~\cite{Golfand:1971iw,Volkov:1973ix,Wess:1974tw,Wess:1974jb,Ferrara:1974pu,Salam:1974ig,Martin:1997ns} with $R$-parity conservation~\cite{Goldberg:1983nd,Ellis:1983ew}.
The supersymmetric partners (superpartners) of the gluon (gluino) and quarks (squarks) are expected to be produced at the LHC if their masses are light enough, while the lightest neutralino\textemdash a mixture of superpartners of the neutral gauge and Higgs bosons\textemdash acts as the DM candidate.
Such scenarios, if realized in nature, would lead at the LHC to signatures with jets with anomalously large transverse momentum and events with significant imbalance in the transverse momentum calculated from all reconstructed particles, \metvec{} with magnitude \met{}.

Although events from multi-jet processes from quantum chromodynamic (QCD) interactions have only little genuine missing transverse momentum, e.g., from neutrinos being produced in electroweak heavy flavour decays, they can exhibit large values of \met{} after the event reconstruction as a result of mis-measured jet momentum.
The yield of such events after the selection cuts is typically expected to be small, but its estimation is challenging because several features of QCD multi-jet events are poorly modelled in Monte Carlo (MC) simulation, owing largely to their governance by non-perturbative interactions. 
These quantities include the production cross section, jet multiplicity, heavy flavour jet multiplicity, and angular relations among jets.
Further challenges arise in the modelling of instrumental effects that can cause \met{}, such as energy loss from non-instrumented or disabled detector regions, jets from calorimetric noise, beam induced background, or cases with pile-up jets being wrongly identified to originate from the hard-scatter interaction.
This motivates the development of data-driven techniques to estimate the multi-jet (or $\gamma+\text{jets}$) background.
Ideally, approaches rely on simulation only for particularly well-modelled quantities, but derive the most important features using real data.
As with any data-driven approach, the method must be robust against possible signal contamination that may bias the prediction.

For searches focusing on large jet multiplicity, the ATLAS collaboration developed a technique to measure the shape of the \met{} significance at low jet multplicity, defined as $S=\met{}/\sqrt{\HT}$, where \HT{} is the scalar sum of all transverse jet momenta in the event.
This shape is used to extrapolate from low to high $S$ values at high jet multiplicity, where the QCD multi-jet background is dominant~\cite{ATLAS:2011nkd}.
For searches at lower jet multiplicity, various techniques have been used:
early searches by CMS used, among other techniques, so-called as \emph{ABCD methods} that exploit that the signal region is defined by requirements on two highly separating observables, e.g., \met{} and the minimal azimuthal angular distance between the leading jets and \metvec{}.
Inverting one or both of those requirements establishes a multi-jet enriched control region that can be used to estimate the multi-jet contribution for the signal region.
The dominant uncertainties arise from the modelling of the correlation of the two chosen observables~\cite{Collaboration:2011ida}.
Another approach used by the ATLAS collaboration is the so-called \emph{jet smearing} method~\cite{ATLAS:2012qgw}.
The main idea is to select well-measured multi-jet events at low $S$ values, so called \emph{seed events}, and then smear the jets of those events with jet response distributions.
The smearing can be performed multiple times to increase the statistics of the produced pseudo data set. For the above methods, large systematic uncertainties often arise as a result of the highly approximate nature of the underlying assumptions of the respective models, or from limited statistics of control regions used by the estimation procedure. 

This document presents the concepts and particulars of the \textit{Rebalance and Smear} (R\&S) method, which is essentially a data-driven, generative QCD model that aims to mitigate issues associated with other estimation methods. A complementary code package is also made available which serves as a generic implementation that interfaces with public tools. Early development of the method was carried out at CMS~\cite{Koay:2011qqa} to measure fake \met{} backgrounds for searches, and later developments were carried out as a component of a series of CMS searches in the all-hadronic channel~\cite{Collaboration:2011ida,Khachatryan:2017rhw,CMS:2017okm}.
A modified version of the method has been used by ATLAS for the search for invisibly decaying Higgs bosons produced in Vector Boson Fusion~\cite{ATLAS:2022yvh}.
The version presented here makes use of the \textsc{Delphes3}~\cite{deFavereau:2013fsa} framework, and the method has been extended to predict backgrounds in final states other than the all-hadronic channel, including channels with one or more photons.

Section \ref{sec:method} gives an introduction to the technical details of the method and describes the various required inputs.
An example use case based on a hypothetical search for pure Higgsino DM is presented in Section \ref{sec:example}.
Section \ref{sec:systematics} provides a discussion of relevant systematic uncertainties along with a list of potential failure modes.
Finally, Section \ref{sec:conclusion} concludes with an outlook about future applications of the method.


%% file: method.tex
\section{Rebalance and Smear methodology} 
\label{sec:method}

The R\&S method is an event-by-event unfolding of, and subsequent application of, the response of the detector to jets.
In the version presented, the unfolding makes use of Bayesian inference to obtain probable values for the true momenta of jets in a seed event from data under the hypothesis that the event originated from a QCD multi-jet process.
This inference is made using a model of the jet response as well as prior knowledge of the distribution of true missing transverse momentum in multi-jet events.
Other implementations of R\&S adjust the momenta of the jets of the seed event within their uncertainties via a kinematic fit in such a way that no missing transverse momentum is present after rebalancing.

The subsequent application of the detector response to the jets amounts to a smearing, or rescaling, of each jet's momentum by a value randomly sampled from the jet response function.
The rebalanced and smeared events constitute a sample that is a proxy for the fake-\met{} background, referred to as the prediction sample.
The prediction sample is then treated like an ordinary MC data set, namely, the analysis is run over the prediction sample to obtain the predicted background yield in a chosen search region. 
The following subsection explains why and in which circumstances the prediction sample can provide a reliable background estimate. 

\subsection{Impact on events with real and fake \met}

When events in an inclusive sample \textit{without} real \met{} are processed through the Rebalance and Smear procedure, the resulting sample conforms well to the original; moreover, the processed events preserve the exponentially-suppressed \met{} tail and jet multiplicity distribution native to the input distribution.
However, when events in an ensemble \textit{with} real \met{} are subjected to the same process, the output conforms to a much more suppressed \met{} tail than the input.
In practice, a typical seed event sample contains a mixture of real and fake-\met{} events, but because the inclusive cross sections of electroweak processes are negligible compared to QCD multi-jet production, the contribution of real \met{} seed events to the rebalanced and smeared \met{} tail is correspondingly negligible.
It follows that the procedure can be safely applied on real data seed samples, and that the output events populating the \met{} tail are a good proxy for the fake-\met{} contribution.

In order for the prediction sample to represent an unbiased estimate of the QCD multi-jet, 
two criteria must be satisfied. 
First, the seed sample must be inclusive with respect to the \met{}, i.e.\ any pre-selection or trigger requirement on the \met{} would deplete the seed sample, causing an underestimate and shape distortion in the prediction sample.
Second, the signal region (SR) must have sufficiently large \met{} such that the contribution of real \met{} events to the prediction is suppressed.
To quantify the remaining contamination, simulated signal events can be processed through the R\&S steps to estimate the rate of the signal's contribution to the background estimate for a given SR.
In typical cases, such contamination is found to be negligible with large \met{} thresholds on the order of $100\,\GeV$.

\subsection{Analyzing the prediction sample}

After rebalancing and smearing the seed events, the prediction sample can be sorted into SRs using the ordinary selection requirements employed by the analysis. 
If the luminosity of the seed sample is the same as that of the final target data, and if an event's jets are smeared just once, then the unweighted counts from the prediction sample in each SR serve as a properly-normalized approximation to the fake \met{} background.
In some cases, appropriate event weights have to be considered, since the seed sample may be selected from data with prescaled triggers.
To gain additional statistics in the prediction sample, each unfolded seed event can be subjected to this smearing process multiple times.
Each smearing iteration operates on the rebalanced event via a unique random sampling of the jet response, yielding a different configuration of jets each time.
This technique can be useful for filling out tails of distributions for signal regions defined by very tight cuts. 
With a sufficiently large number of smearing iterations, a non-zero prediction in each SR is reachable.
An event weight of $1/n$ restores the prediction to the correct scale, where $n$ is the total number of smearing iterations applied to the corresponding seed event.
The assignment of statistical uncertainty to a prediction sample in which seeds have been re-used in this manner is carried out using a bootstrapping technique.
Here, a number of small subsets, advisably order 10 or larger, are drawn from the complete seed sample, and the prediction is made once per subset.
A weight is applied to the prediction from each subset equal to the ratio of total number of seed events to the number of seeds in the subset.
From the ensemble of predictions, the standard deviation taken to represent the statistical uncertainty in the prediction.

\subsection{Event samples}

This background estimation implementation is based on simulated $pp$ collision events with center-of-mass energy of $14\,\TeV{}$ generated and hadronized using \textsc{Pythia8.1}~\cite{Sjostrand:2007gs}.
For these samples, the production flag sets \texttt{HardQCD:all}, \texttt{SUSY:gg2squarkantisquark}, \texttt{ffbar2W}, and \texttt{ffbar2gmZ} are used.
After hadronization and simulated parton showering, events are processed with the detector simulation program \textsc{Delphes3}~\cite{deFavereau:2013fsa} using a CMS-like detector geometry and resolution.
Jets are clustered using the anti-$k_{\text{T}}$ jet algorithm implemented within \textsc{FastJet}~\cite{Cacciari:2011ma} with a jet cone size parameter of 0.4.
Generator-level jets are clustered from the generated particles with \textsc{Pythia} status=1, neglecting neutrinos.
Reco-level jets are obtained by clustering all reconstructed final state particles identified by \textsc{Delphes3}, which includes electrons, muons, photons, and hadrons.

\subsection{QCD jet and event model}

\subsubsection{Likelihood of response}

\begin{figure}[tbh]
\centering
\subfloat[]{
\includegraphics[width=0.45\linewidth]{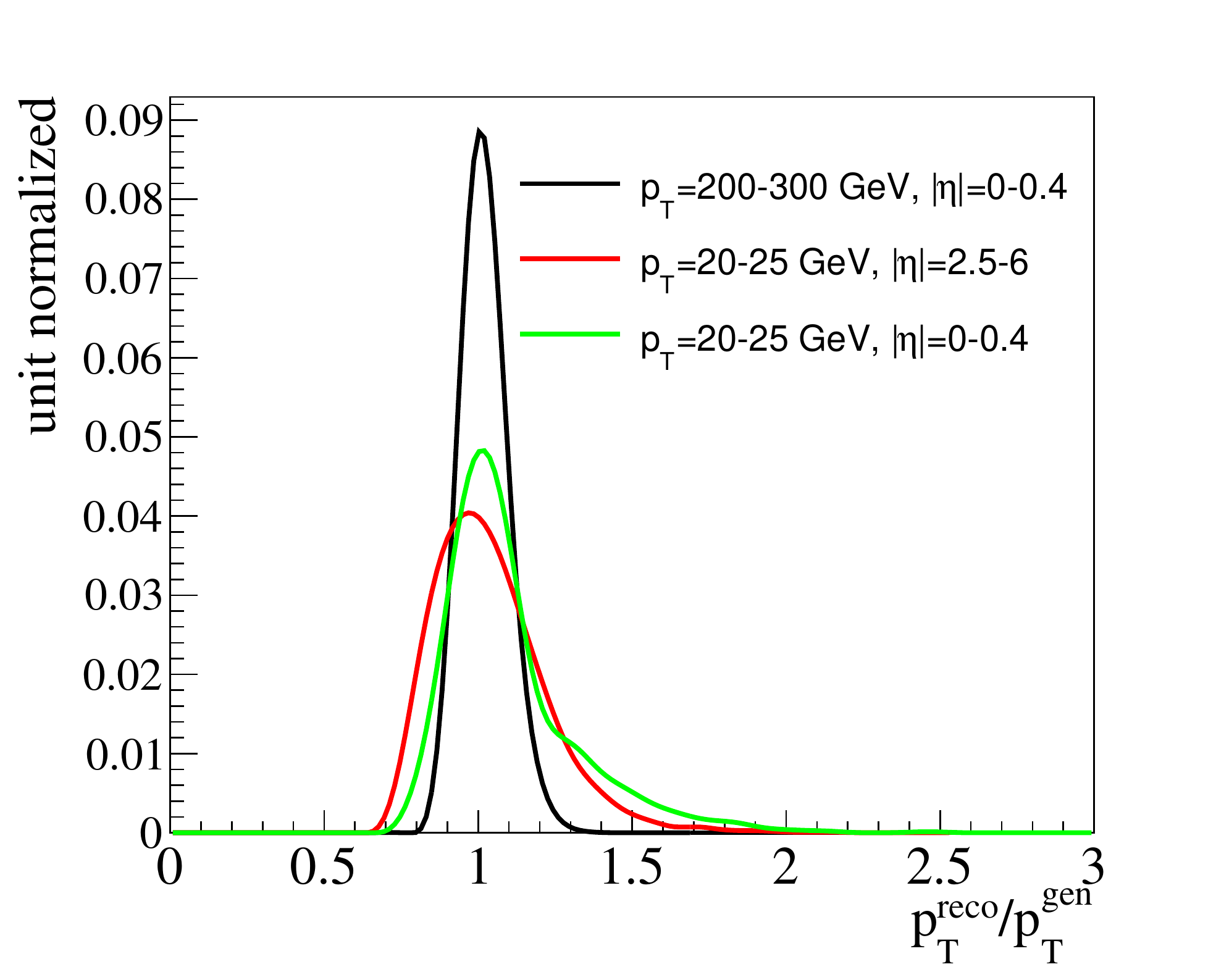}
}\\
\subfloat[]{
\includegraphics[width=0.45\linewidth]{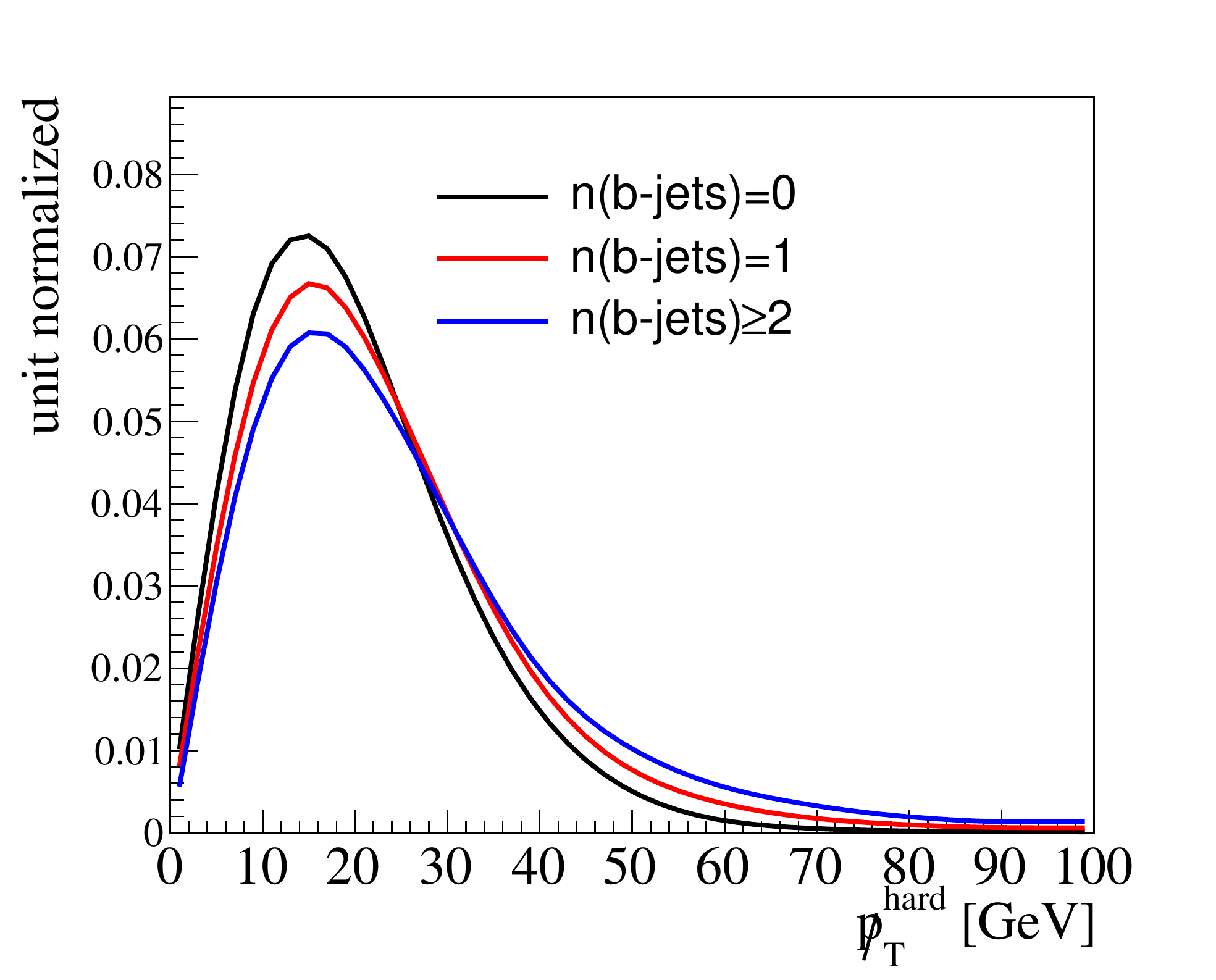}
}
\subfloat[]{
\includegraphics[width=0.45\linewidth]{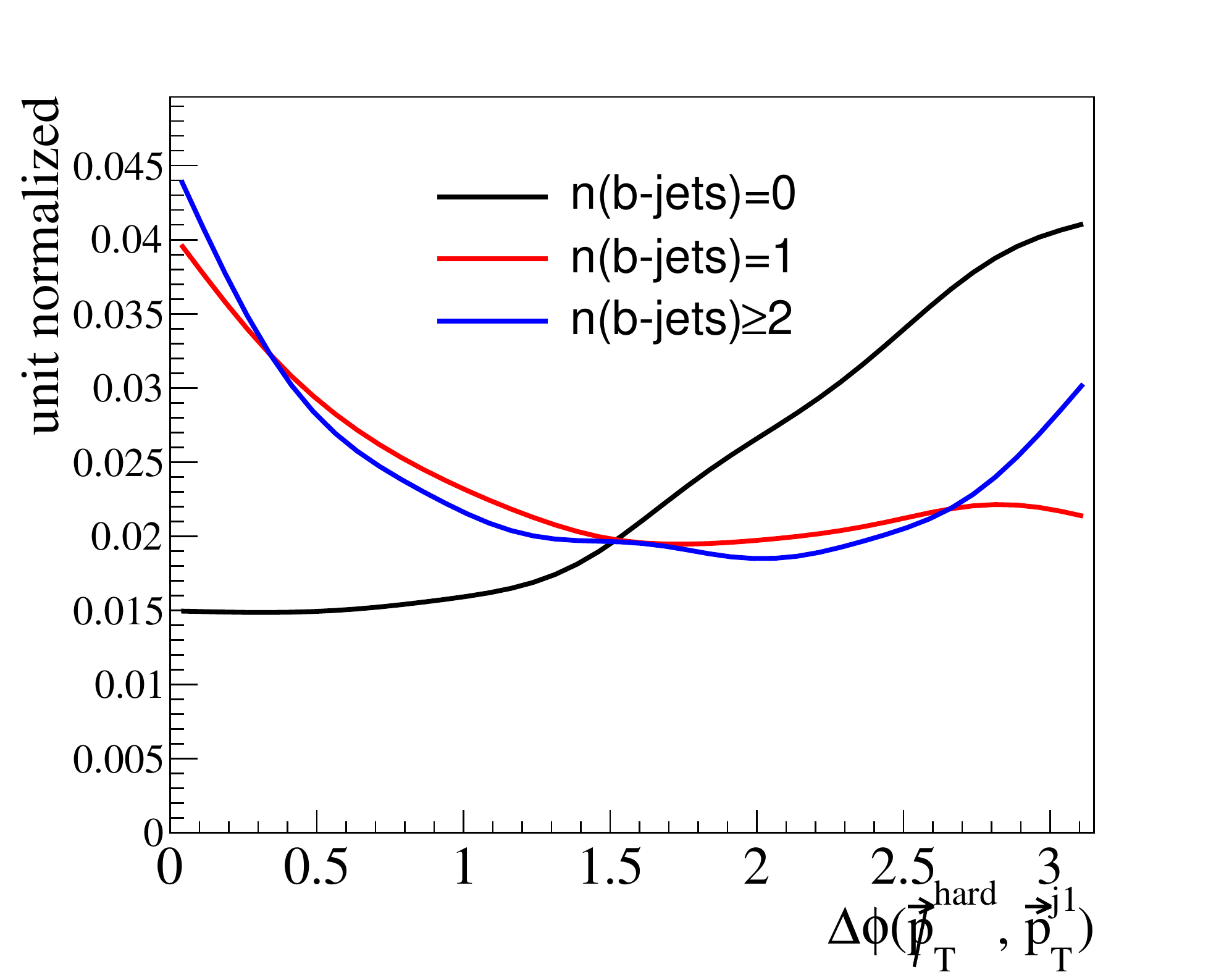}
}
\caption{Jet response implemented for the \textsc{Delphes} simulation for three different choices of ranges for $p_{\text{T}}$ and $\eta$ (a), used as input to the likelihood and as smearing templates. PDF of the generator-level \methard{} (b) and azimuthal angle with respect to \methard{} (c) of in simulated QCD multi-jet events, factors in the rebalancing prior.}
\label{fig:SmearEx}
\end{figure}

Both the R\&S steps rely on a supplied model for the jet energy response distribution. 
In the presented implementation, the response distribution is derived from simulated QCD multi-jet events with the response defined for a matched pair of reco-level and generator-level jet as the ratio of the reco-level jet energy to the generator-level jet energy.
Further, it is assumed that the directions of the $n_{j}$ jets in a given event are measured with ideal accuracy, which is well motivated by comparing the good angular resolution of the detectors to their larger energy resolution:
\begin{equation}
\boldsymbol{J}_{\text{reco}} = \hat{C}\boldsymbol{J}_{\text{true}},
\end{equation}
where $\boldsymbol{J}_{\text{reco}}$ and $\boldsymbol{J}_{\text{true}}$ are respectively length-$n_{j}$ vectors of the reconstructed and true jet four-vectors, $\hat{C}$ is a diagonal matrix of jet mis-measurement scale factors $(c_1,c_2,...,c_n)$.
The probability for a measurement to correspond to a particular $c_{i}$ given a true momentum $p_{i,\text{true}}$ is the single-jet likelihood
\begin{equation}
L_{i}\equiv P(p_{i,\text{reco}}/p_{i,\text{true}}|\ p_{i,\text{true}}) = P(c_{i}|\ p_{i,\text{true}}),
\label{eq:JetmomentumLikelihood}
\end{equation}
also referred to as the probability density function (PDF) of the jet response. The response is binned in ranges of $p_{\text{T}}$ and $\eta$, and a few examples are shown in Figure~\ref{fig:SmearEx} (left).
The response function for a single jet is obtained by linearly interpolating between binned PDF functions.
A reasonably accurate model of the jet response is key, as the associated systematic uncertainty is often the predominant one.

The choice to neglect the energy of neutrinos when clustering generator-level jets ensures that the response is approximately the same for different jet flavours.
This allows a single set of jet response templates to be used rather than multiple sets for jet flavour categories.
In a scenario with data and not simulated events, either the shape of the jet response should be extracted directly from the data, or the width and scale of the distributions derived from simulated data or by other means should be validated and corrected for any mis-modelling.
The association of generator-level and reco-level jets used for the construction of the likelihood is made via a matching criterion given by
\begin{equation}
\Delta R(\boldsymbol{p}_{\text{reco}},\boldsymbol{p}_{\text{true}})<0.4.
\end{equation}	
An isolation criterion of $p_{\text{T}}^{\text{sum}}/p_{\text{T}}<0.01$ is applied to each selected generator-level (reco-level) jet, where 
\begin{equation}
p_{\text{T}}^{\text{sum}} = \sum_{i=1}^{n_{\text{jets}}}(p_{\text{T}})_{i}
\end{equation}
is the sum over generator-level (reco-level) jets within $\Delta R<0.5$ of the selected jet, excluding the selected jet.
This ensures a clean definition of the jet energy response with no artificial effects, e.g.\ from one generator-level jet being reconstructed as two reco-level jets, which would lead to very small response values.

\subsubsection{Prior probability distribution for \met}

Rebalancing additionally relies on a provided PDF of the true missing transverse momentum of QCD multi-jet events, as well as the azimuthal angle $\Delta\phi(\slashed{\boldsymbol{p}}_{\text{T}}^{\text{hard}}, \boldsymbol{p}_{\text{T}}^{j1})$ between the leading jet and the true \met{} (prior).
These two PDFs are treated as separable, and are taken from generator-level events.
They are parametrized in categories of the multiplicity of $b$-tagged jets, where the $b$-tagging emulation from \textsc{Delphes3} is used. 
The distribution of the generator-level \met{} is shown in Figure~\ref{fig:SmearEx} (right). These PDFs give expression to the prior probability for an arrangement of jets in a given event as

\begin{equation}
\pi(\boldsymbol{J}_{\text{true}}) = P[\Delta\phi_{\metvechard,\vec p_{\text{T}}^1}(\boldsymbol{J}_{\text{true}})]\cdot \pi_0(\boldsymbol{J}_{\text{true}}),
\label{eq:prior1}
\end{equation}
where $\pi_0(\boldsymbol{J}_{\text{true}})$ is an initial prior (``ur"-prior) on the parton jet four-vectors, taken to be a constant so as not to impose a bias on the $p_{T}$ spectrum of jets.

The quantity that constrains the posterior density the most is the prior density for the true \met{}.
For QCD multi-jet events, the true \met{} is characteristically small since a finite value arises only from low-momentum neutrinos appearing in the decays of heavy-flavour hadrons in jets, or from particles falling outside of the acceptance of the selection used when computing the \met{}. 
At high instantaneous luminosities additional jets from pile-up interactions are present, typically with small \pt.
Those additional jets deteriorate the \met{} resolution, and therefore, to stabilize the \met{} against effects from pile-up, it is practical to consider a proxy for the \met{} referred to here as hard \met{} or \methard{}, defined as
\begin{equation}
\methard = \left\vert - \sum_{i}^{n_{\text{obj}}} (\boldsymbol{p}_{\text{T}i}\cdot\Theta(p_{\text{T}i}-30\,\GeV)\cdot\Theta(5-|\eta_{i}|) \right\vert.
\label{eq:hardmet}
\end{equation}
The Heaviside functions $\Theta$ serve to remove many jets with low $\pt$ originating from pileup interactions, and reduce the number of dynamic objects contributing to the sum.
The hard \met{} is highly correlated to the traditional \met{} and is suitable for DM searches. At a later stage, this choice results in a more manageable number of free parameters present in the rebalance step of the procedure.
For the remainder of this document we adopt the \methard{} in place of \met{} in all cases. 

To prevent biases that originate from jets migrating during the R\&S phases across the $p_{\text{T}}$ threshold of the considered analysis, i.e.\ $30\,\GeV$, a looser \pt{} criterion, i.e.\ $15\,\GeV$, is used for the jets to enter the R\&S procedure.

\subsection{Rebalance procedure}

The goal is to estimate the probable configuration of jets $\boldsymbol{J}_{\text{true}}$ for each event, given that a particular set of measured jet momenta $\boldsymbol{J}_{\text{reco}}$ has been made, that a well-defined model for the jet response function is available, and that some prior knowledge of the true \methard{} of QCD multi-jet events.
In this context, Bayes' theorem states
\begin{equation}
P(\boldsymbol{J}_{\text{true}}|\boldsymbol{J}_{\text{reco}}) = P(\boldsymbol{J}_{\text{reco}}|\boldsymbol{J}_{\text{true}})\cdot \pi(\boldsymbol{J}_{\text{true}}),
\label{eq:posterior1}
\end{equation}
where $\pi(\boldsymbol{J}_{\text{true}})$ is the $n$-dimensional prior probability density for the true jet collection, which encodes the low-\met{} constraint as well as information about the angle of azimuthal incidence between the leading jet and the vectorial \methard.
Treating the jets as independent from each other and assuming the directionality of jet momenta is measured precisely allows the likelihood to be factorized and written as
\begin{equation}
\begin{split}
P(\boldsymbol{J}_{\text{reco}}|\boldsymbol{J}_{\text{true}}) =  \prod_{i=1}^{n_{j}}L_{i} = \prod_{i=1}^{n_{j}}P(c_{i}\ |\ \boldsymbol{p}_{i,\text{true}}).\\
\end{split}
\label{eq:likelihood1}
\end{equation}
Combining Eqs. \ref{eq:posterior1}, \ref{eq:likelihood1} gives
\begin{equation}
P(\boldsymbol{J}_{\text{true}}|\boldsymbol{J}_{\text{reco}}) =
\prod_{i=1}^{n_{j}}P(p_{i,\text{reco}}|\ c_{i})\cdot P[\methard(\boldsymbol{J}_{\text{true}})]\cdot P[\Delta\phi_{\metvechard,\vec p_{\text{T}}^1}(\boldsymbol{J}_{\text{true}})]\cdot \pi_0(\boldsymbol{J}_{\text{true}}).
\end{equation}

This posterior density is maximized for each seed event, which leads to a single best-fit configuration $\boldsymbol{J}_{\text{true}}^{*}$, corresponding to a vector of best-fit $c_{i}$ values for each event.
The $p_{T}$ values of jets are then smeared according to the jet response, and events are analyzed to form the prediction. The following are choices made in this implementation regarding the rebalancing procedure that have been tested and found to work adequately. 

Before calling the maximization routine, a suitable initialization of the $c_{i}$ values is searched for that helps to ensure convergence by placing the initial density on the smooth, well-defined portion of the posterior PDF.
For the initialization, it is checked whether a reasonable value of hard $\slashed{p}_{\text{T}}$ can be obtained via the scaling of a single jet \pt, where reasonable means that $\methard(\boldsymbol{J^{*}}_{\text{true}})$ resides somewhat near the maximum value of the PDF, where $J^{*}$ is a vector of the modified (initialization) values of the jets.
An initialization target \methard{} is identified for a given event as
\begin{equation}
\text{target }\methard = \text{max}\,[30\,\GeV,\,\text{min}[90\,\GeV,\,\HT/3]],
\end{equation}
where \HT{} is the magnitude of the scalar sum of all jets in the event. It is then checked whether any single jet can have its magnitude re-scaled such that the re-calculated hard $\slashed{p}_{\text{T}}$ assumes the above target value.
If such a solution exists, which results in $|c_{i}^{\text{init}}-1|<0.8$ for jet $i$, the corresponding configuration is chosen for the initialization, and other jets are initialized to their unscaled values.  

The maximization is performed with the \textsc{ROOT} \texttt{TMinuit} package in order to rebalance the event, where each $c_{i}$ is allowed to float with a step size 0.05.
The PDF's for the likelihood (and prior) are linearly interpolated between the bins of $p_{\text{T}}$ and $\eta$ ($H_{\text{T}}$).
This ensures a mostly smooth gradient for the posterior density.
A discontinuity can however occur if the multiplicity of $b$-tagged jets changes during the rebalancing procedure, but such cases are found to be rare.

\begin{figure}[tbh]
\centering
\subfloat[]{
\includegraphics[width=0.5\linewidth]{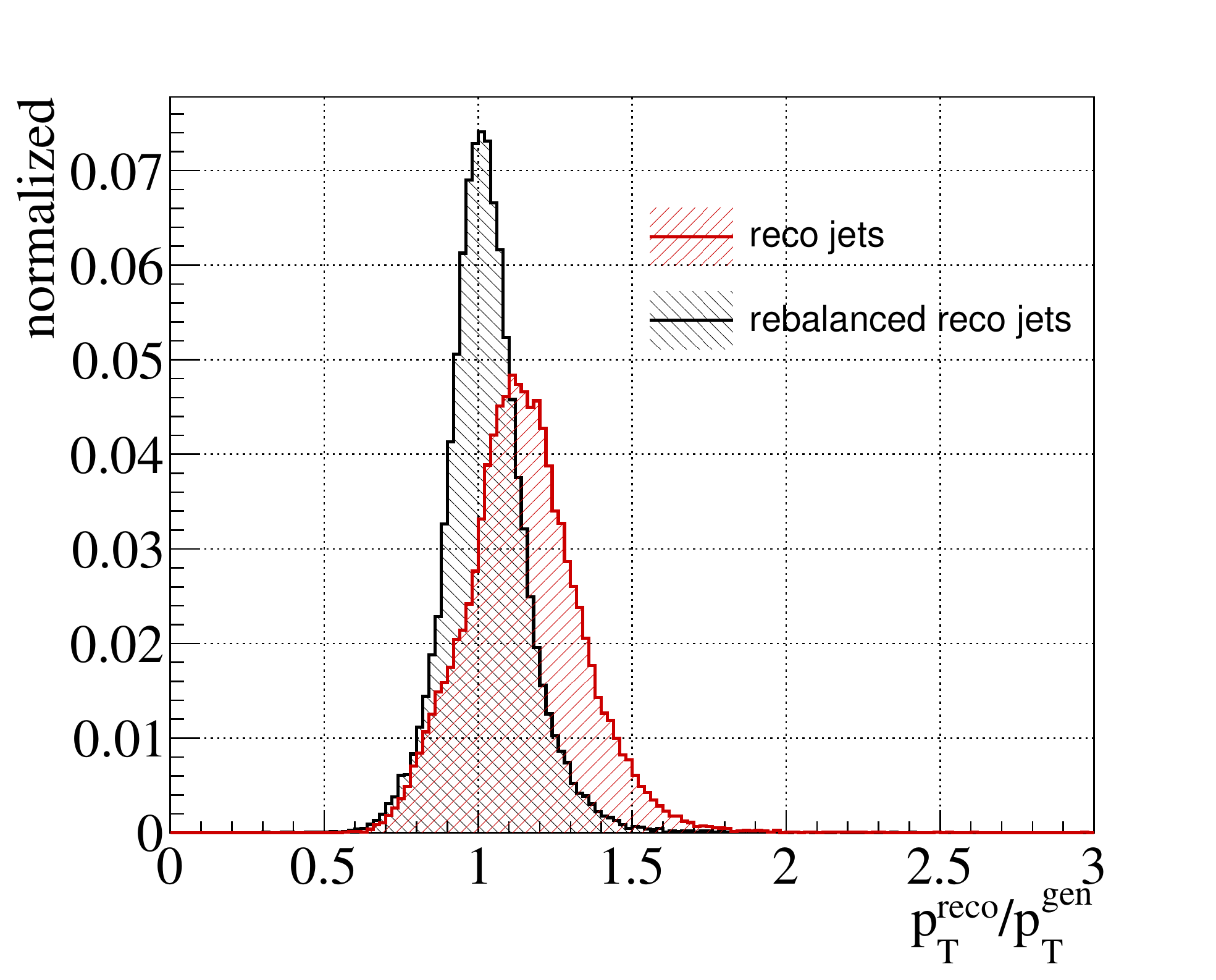}
}
\subfloat[]{
\includegraphics[width=0.5\linewidth]{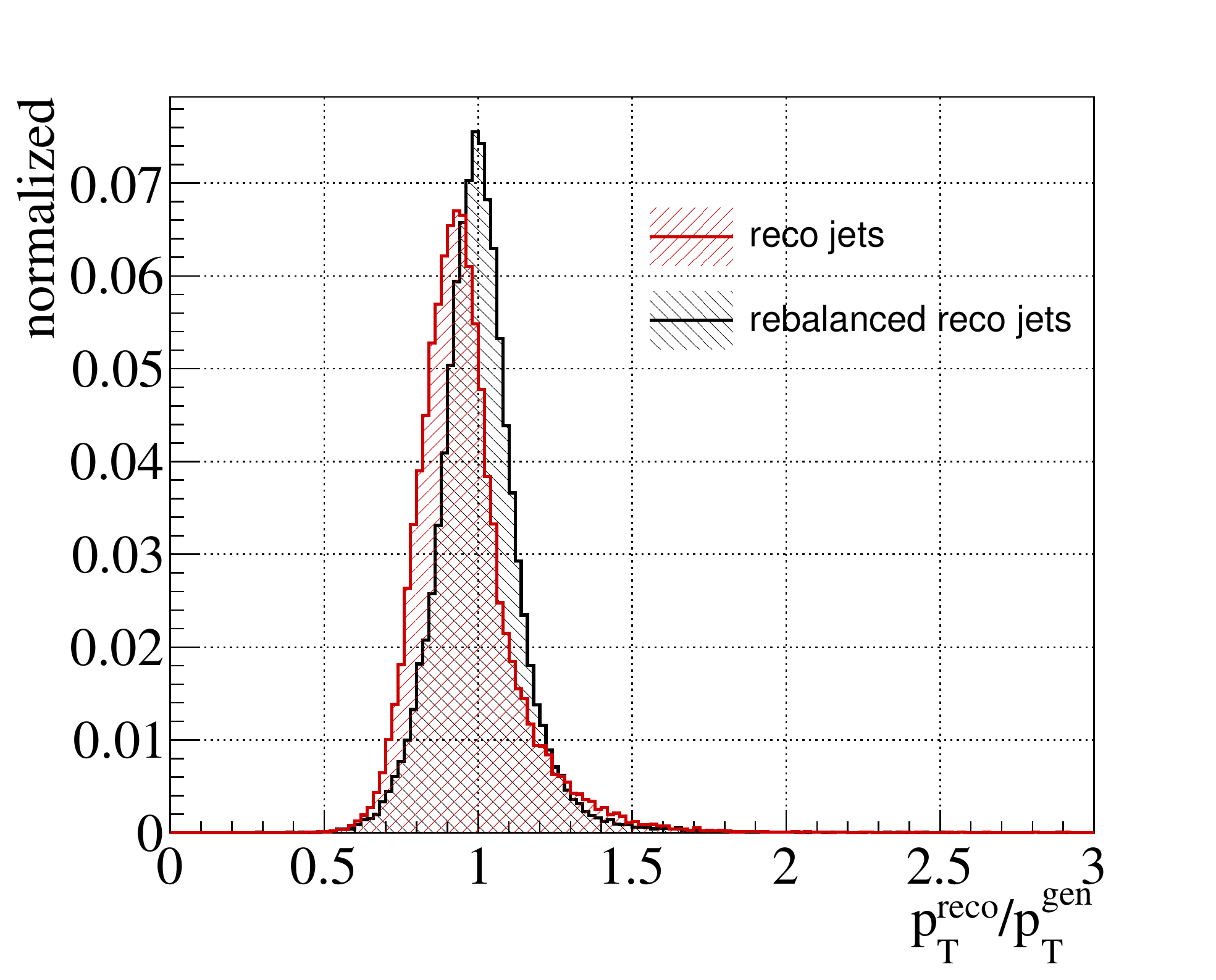}
}
\caption{Distributions of the jet response of in a sample of simulated QCD multi-jet events before (red) and after (black) rebalancing for the leading (a) and sub-leading (b) jet in events with large missing transverse momentum ($\slashed{p}_{\text{T}}>120$ GeV).
The response distributions are constructed as histograms of the ratio of the reconstructed or rebalanced jet $p_\text{T}$ to the generator-level $p_\text{T}$.
Events are selected in a signal region with large, which enhances the prevalence of jet mis-measurements.}
\label{fig:randsBeforeAfter}
\end{figure}

Several of the above choices have been developed empirically.
The convergence rate for simulated QCD multi-jet events is found to exceed 99\% overall.
Events for which the maximization fails to converge are discarded, as well as any events which fail to arrive at a sufficiently small value of the rebalanced \methard, in the case of the example presented, of $100\,\GeV$.
As a demonstration that the \met{} prior meaningfully restores truth-level information of an event with one or more poorly-reconstructed jets, a test is performed on a sample of simulated events passing a baseline selection that requires hard $\methard>200\,\GeV{}$ and $n_{j}\geq2$.
The distribution of the \pt{} ratio of the reconstructed and matched generator-level jets, i.e.\ the jet response distributions, are compared for the leading and sub-leading jets in Figure \ref{fig:randsBeforeAfter}, before and after rebalancing.
While the original response distributions exhibit evidence of severe mis-measurement, those of the rebalanced jets are found to peak more narrowly at 1. 

\subsection{Events with well-measured objects}

Objects who's momentum and and energy are expected to be very well-measured compared to jets, e.g., electrons, photons, and muons over common kinematic ranges, are not modified during rebalancing or smearing.
In a generic way, an appropriate resolution model can be applied to these objects to be used for rebalancing and smearing, but it is generally found that  fixing such objects' momentum at their measured values suffices to provide a description without visibly biasing the background estimate. 

%% file: example.tex
\section{Example: search for Higgsino dark matter}
\label{sec:example}

The CMS and ATLAS experiments have excluded DM masses up to a maximum of around $1$\textendash$2\, \TeV$ in gluino models and to several hundred$\, \GeV$ in certain models with squark pair production~\cite{Sirunyan:2019ctn,Aad:2020aze}.

Despite huge improvements in sensitivity during Run-1 and Run-2 of data taking, no clear signal of DM production has been observed yet.
However, there remain prominent regions of still-unexplored phase space with the potential to provide a DM candidate, which would have been produced abundantly in the data sets already collected by CMS and ATLAS.
Canonical examples are so-called compressed regions of simplified models~\cite{Alwall:2008ve,Alwall:2008ag,Alves:2011wf}, where a DM candidate particle and an associated heavier particle have masses near to each other, with a mass difference in the range of a few or several hundred GeV.
For such scenarios, the momentum of the visible decay products is limited, resulting in low transverse and missing transverse momentum and thus a loss of acceptance in the analyze search regions. 

An example of a non-excluded compressed region a model with a kinematically accessible squark and a Higgsino dark matter candidate, which means that the lightest neutralino is almost a pure superpartner of the Higgs boson.
Higgsinos can fully account for the dark matter relic density $\Omega_{h}$ if they have a mass of around $1.1\,\TeV$~\cite{Delgado:2020url}, which puts them slightly beyond the boundary of limits established by searches for squarks (or gluinos) decaying into DM at the Run-2 LHC\textemdash for example, the searches by CMS~\cite{Sirunyan:2019ctn}, and likewise by ATLAS~\cite{Aad:2020aze}, indicate the sensitivity boundary falling short of this LSP mass value despite large signal production cross sections. 

In general, the sensitivity diminishes as the model spectrum becomes more compressed because signal events exhibit more background-like characteristics, most importantly, they exhibit a more rapidly falling \met{} spectrum.
To target these scenarios, signal regions with low thresholds on the \met{} must be employed, but these regions are characteristically overrun by events from the so-called fake-\met{} background.
This background refers to SM events free of high-$p_{T}$ neutrinos that nonetheless have a large transverse momentum imbalance among the reconstructed objects due to detector resolution effects.
The most prominent example of this background arises from QCD multi-jet production, which by far accounts for the majority of events produced at the LHC.
Depending on the selection, it is a significant background to searches in the all-hadronic channel, as well as channels with one or more photon.

As proof of principle of the R\&S method, a possible search for evidence of Higgsino DM is explored with simulated events.
The characteristic small mass difference of the lightest and next-to-lightest supersymmetric particles leads to final states with not too large values of \met{} and consequently the sensitivity of the search improved if the QCD multi-jet background is reliably estimated.
The data-driven R\&S method is applied to the simulated events to predict the QCD multi-jet yield for a signal enriched selection.
The performance of the method can be quantified by a comparison of the R\&S prediction with the yields obtained directly from the simulated samples.

\subsection{Signal model attributes}

\begin{figure}[tbh]
\centering
\includegraphics[width=0.45\linewidth]{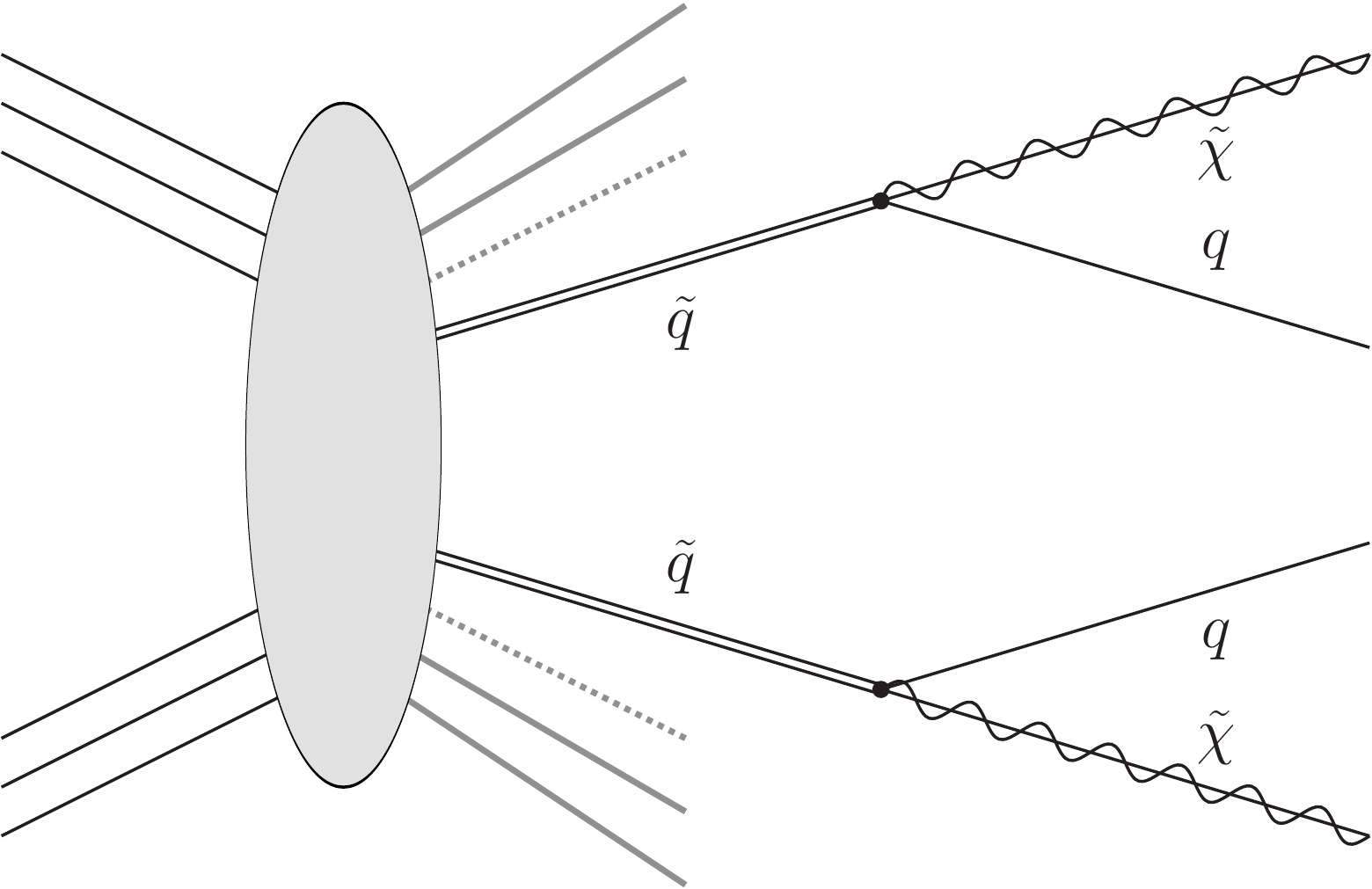}
\caption{Schematic diagram representing a simplified T2qq model. Each produced heavy particle, i.e.\ the two squarks in this example, decay into one quark and an undetectable particle, i.e.\ the neutralino.}
\label{fig:T2qq}
\end{figure}

The model for the signal process is a simplified model for direct squark production, which is referred to as the T2qq model and which is commonly used in CMS and ATLAS papers for interpreting searches for supersymmetry
A schematic diagram of the squark production process is shown in Fig.~\ref{fig:T2qq}.
Specifically, signal events feature the pair production a quark/anti-squark pair, where each (anti-) squark decays into an (anti-) quark and a neutralino ($\tilde{q}\rightarrow q\tilde{\chi_{1}}^{0}$), the neutralino being the stable dark matter candidate.
In the present work, the more optimistic version of the model is assumed, where the squark has an 8-fold degeneracy among the first and second generation flavour states, each having a left- and right-handed squark.
Such scenarios appear in constrained SUSY models such as the cMSSM~\cite{Kane:1993td}.

As mentioned, the Higgsino LSP $\tilde{\chi}_{1}^{0}$ must have a mass of around 1.1 TeV in order to fully explain the DM relic density.
Therefore, a benchmark signal model point is chosen corresponding to a Higgsino mass of 1.1 TeV and a squark mass of 1.15 TeV, a set of parameters that remains not excluded by searches from CMS and ATLAS~\cite{Sirunyan:2019ctn,Aad:2020aze}.
This configuration occupies the compressed region, where the phase space of the visible squark decay products is limited due to the small mass difference between the squark and neutralino.

\subsection{Event selection}

\subsubsection{Pre-selection and classifier training}

An event pre-selection is defined to be consistent with LHC searches and with standard \met{} triggers.
Events are required to satisfy

\begin{itemize}
\item \methard{}$>120$ GeV, computed with jets with $p_{T}>30$ GeV and $|\eta|<5.0$;
\item $n_{\text{jets}}>0$, counting jets with $p_{T}>30$ GeV and $|\eta|<2.4$;
\item $n_{\text{b-jets}}=0$, where a b-tagging efficiency of $\sim$80\% is employed;
\item $n_{\text{electron}}=n_{\text{muon}}=0$;
\item and $H_{T}>$ \methard{}.
\end{itemize}
The pre-selection is applied on all simulated events, as well as on all rebalanced and smeared simulated events.
Note that events are and must be rebalanced and smeared before the application of the pre-selection, given the looser object selection used during the R\&S procedure.
Each rebalanced event is copied and independently smeared 100 times to increase the statistical precision of the prediction in the baseline and signal regions.

A multivariate classifier (BDT) is trained using the simulated signal and background events passing the pre-selection using the ROOT TMVA package~\cite{Hocker:2007ht}.
The classifier is trained to discriminate between T2qq events (signal) and electroweak boson events (background).
Note that the rebalanced and smeared events can also be used as input to the training, but in the idealized detector it is found that the signal events most significant kinematic overlap with the electroweak background. A comprehensive set of event kinematical observables is used as input to the BDT, including the hard \methard{}, the $H_{T}$, as well as the $p_T$, $\eta$, and $\phi$ of the four highest-$p_{T}$ jets.
For events with fewer than four jets, the inputs coding information from the non-existent jets are set to 0.
The azimuthal coordinates of all jets $\phi$ are taken with respect to the $\boldsymbol{\slashed{p}}_{\text{T}}$ vector.
The BDT is chosen to have 200 trees and a maximum depth of 4.
No over-training is observed in comparisons between the training samples and a statistically independent validation sample. 

\subsubsection{Baseline and signal region selection}

The baseline selection is defined as the set of events passing the pre-selection, as well as passing a tighter cut on the hard \methard{} of 250 GeV.
This requirement insures that the events are fully efficient with respect to typical \methard{} triggers employed by CMS and ATLAS, and serves to improve the signal-to-background ratio. 

\begin{figure}[tbh]
\centering
\subfloat[]{
\includegraphics[width=0.45\linewidth]{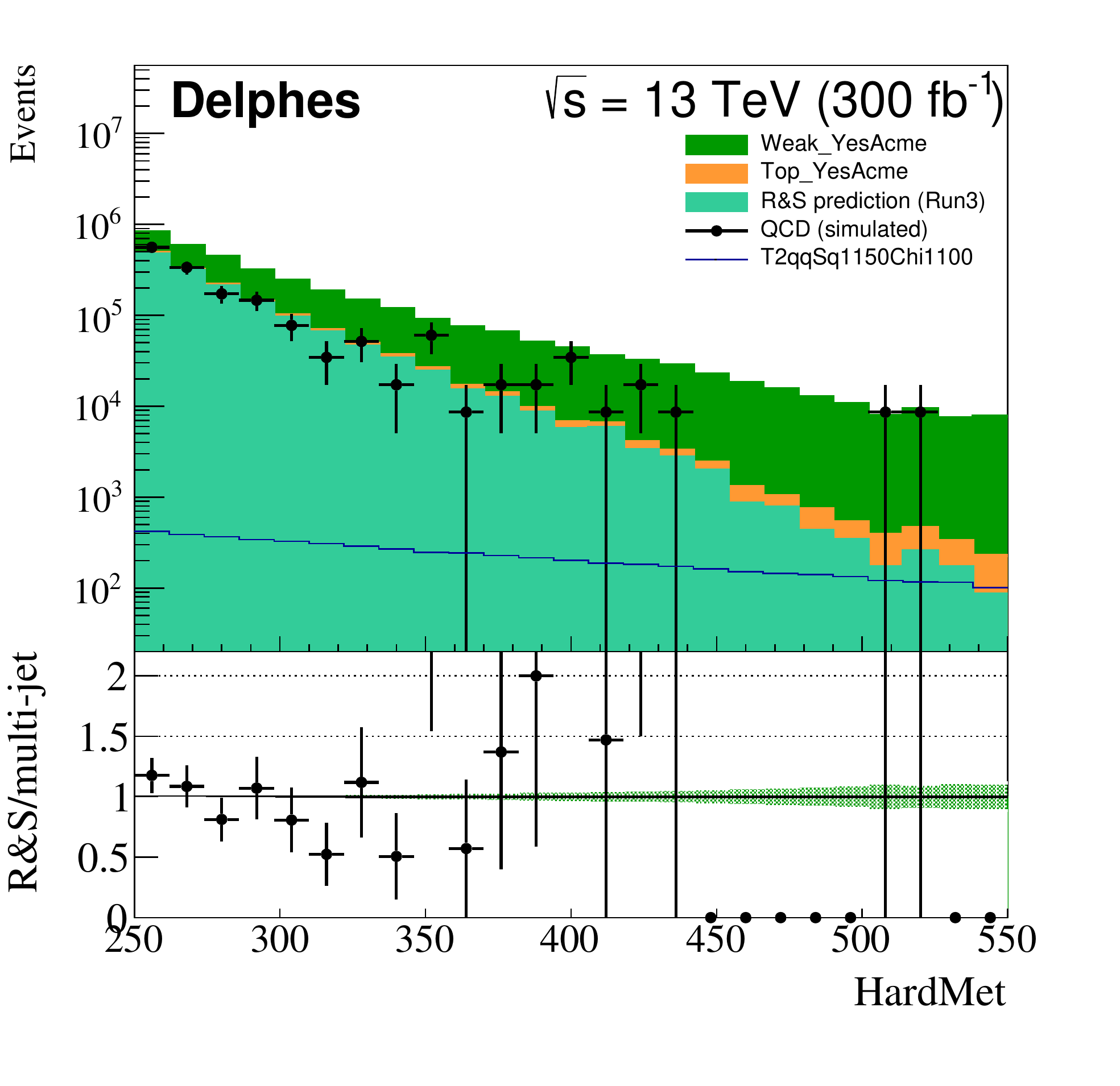}
}
\subfloat[]{
\includegraphics[width=0.45\linewidth]{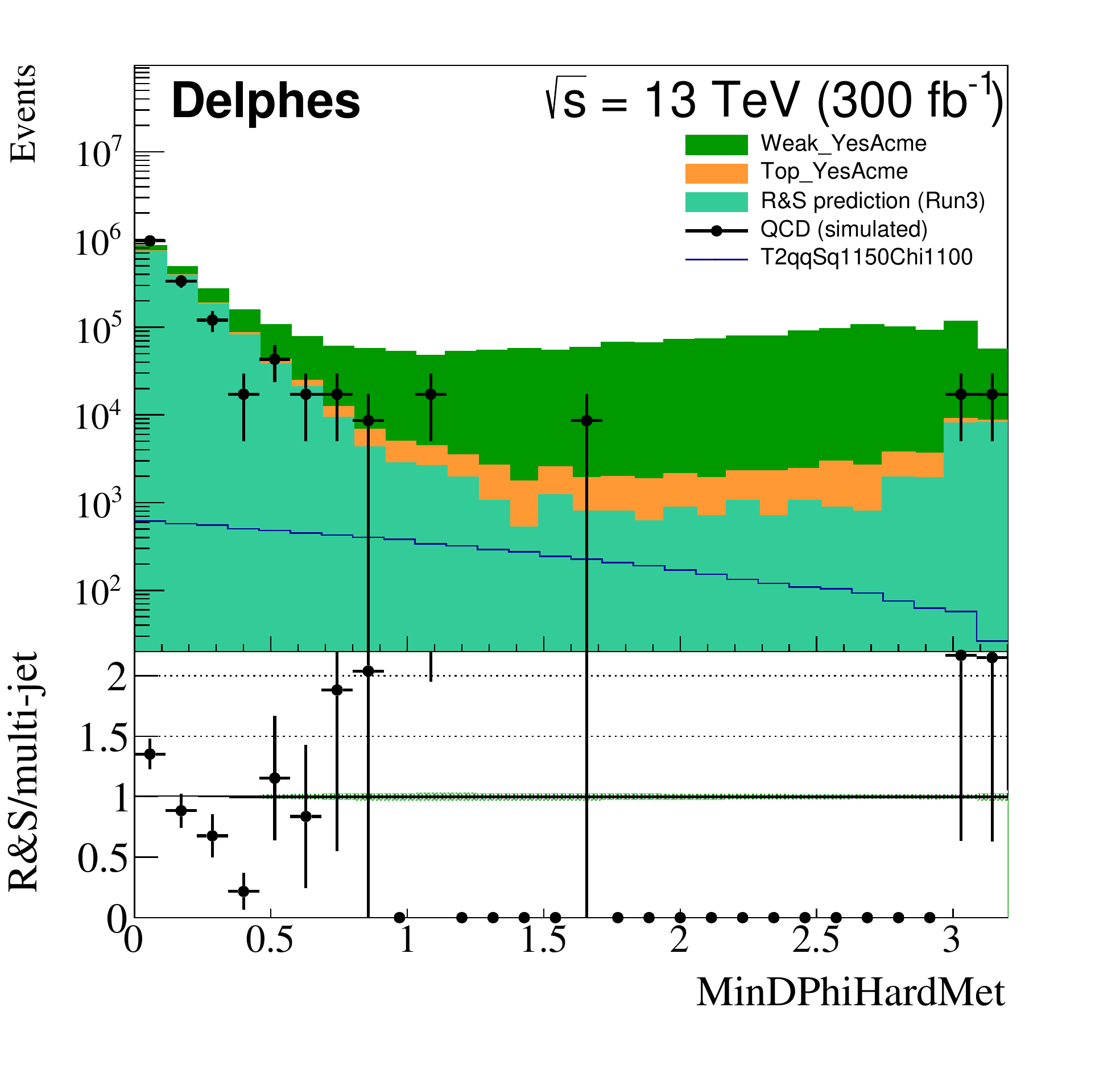}
}\\
\subfloat[]{
\includegraphics[width=0.45\linewidth]{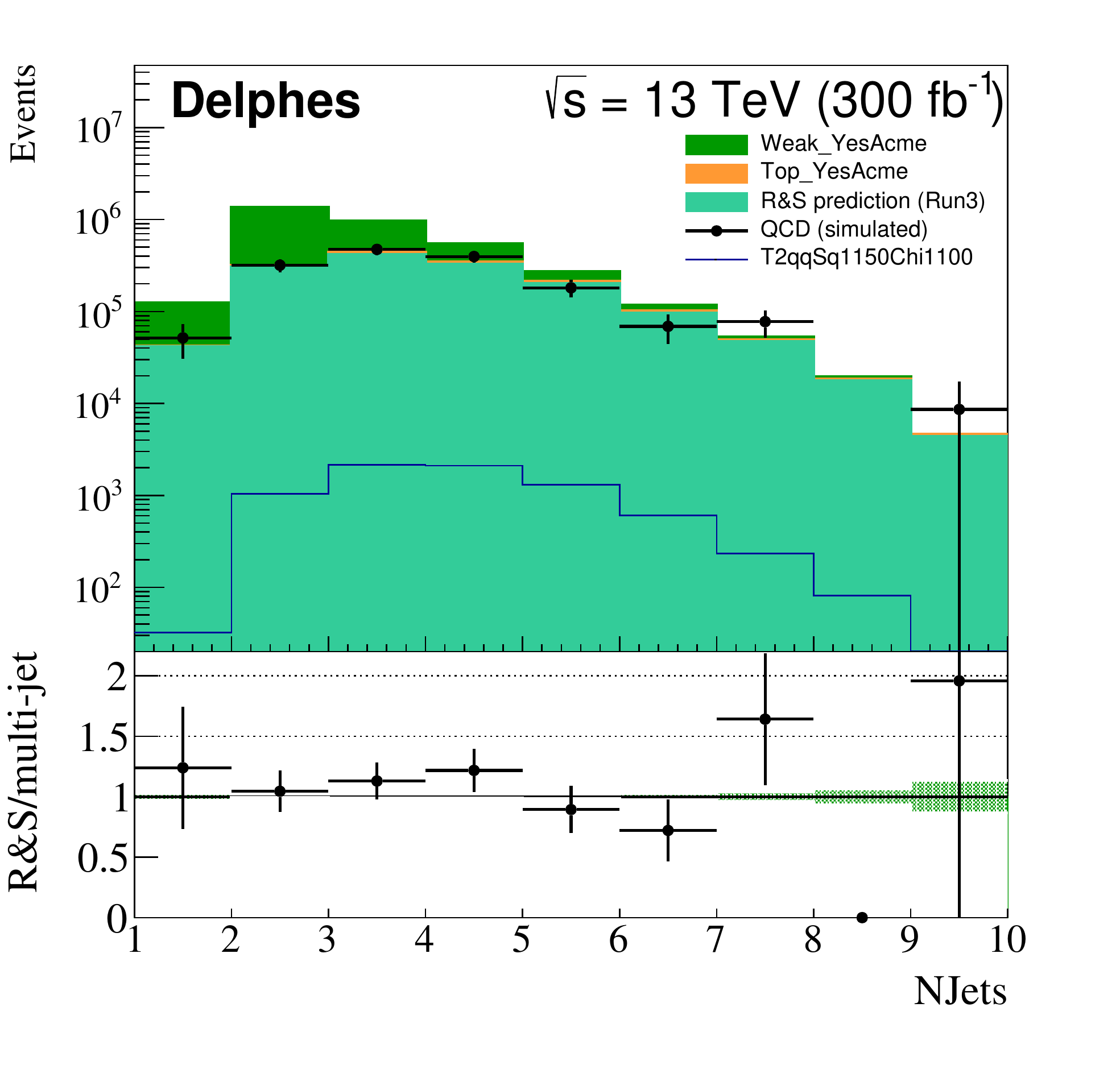}
}
\subfloat[]{
\includegraphics[width=0.45\linewidth]{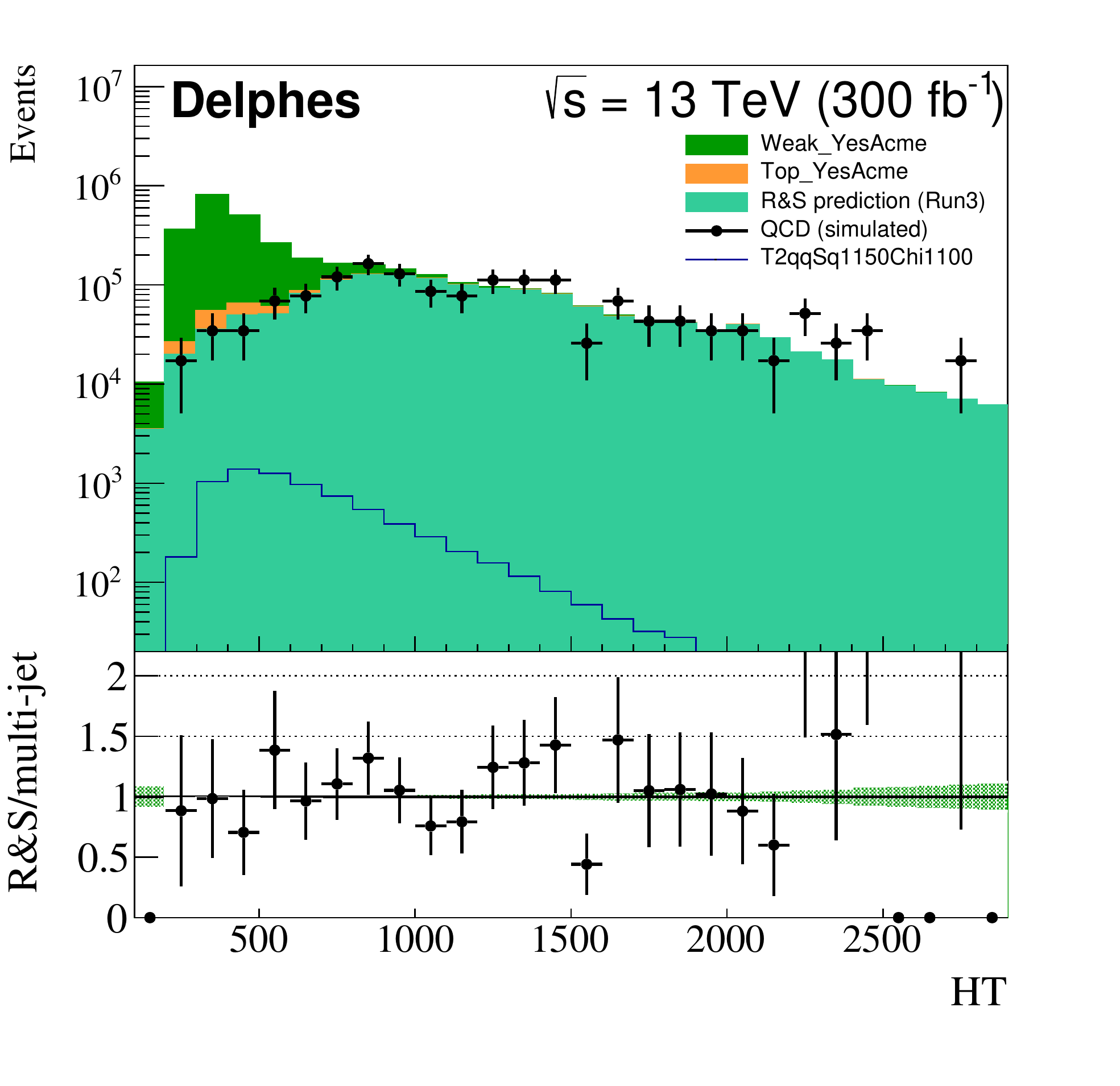}
}
\caption{Distributions of the analysis observables for signal and background events after the event pre-selection. The QCD background obtained directly from simulation is shown as black dots, while the QCD prediction obtained from the R\&S method, is shown in turquoise, and the lower panel indicates the ratio of the two. The QCD prediction is seen to give reasonable agreement with the simulated QCD.}
\label{fig:baseline}
\end{figure}

Distributions of the signal and background events passing the baseline selection are shown in Figure \ref{fig:baseline}.
The stack of shaded histograms show the background predictions where the QCD multi-jet estimate is given as the distribution of the rebalanced and smeared data set, and the non-QCD background is taken directly from simulation.
The QCD distribution taken directly from simulation is drawn as black dots with error bars and is compared bin-by-bin to the R\&S prediction in the ratio panel.
The QCD prediction is seen to give reasonable agreement with the simulated QCD, with no evidence of a statistically-significant deviation beyond about 20\%. 

\begin{figure}[tbh]
\centering
\subfloat[]{
\includegraphics[width=0.45\linewidth]{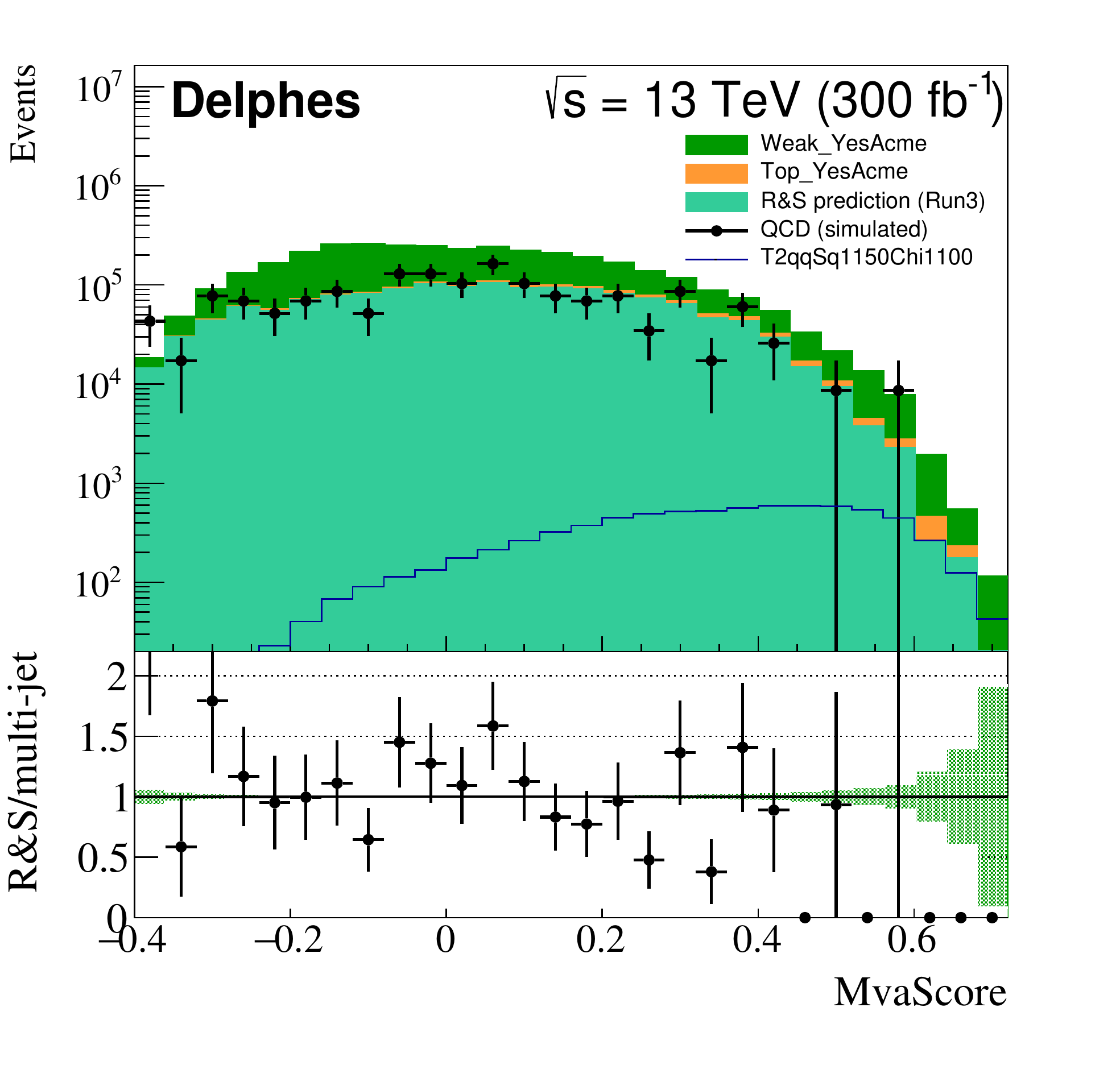}
}
\subfloat[]{
\includegraphics[width=0.45\linewidth]{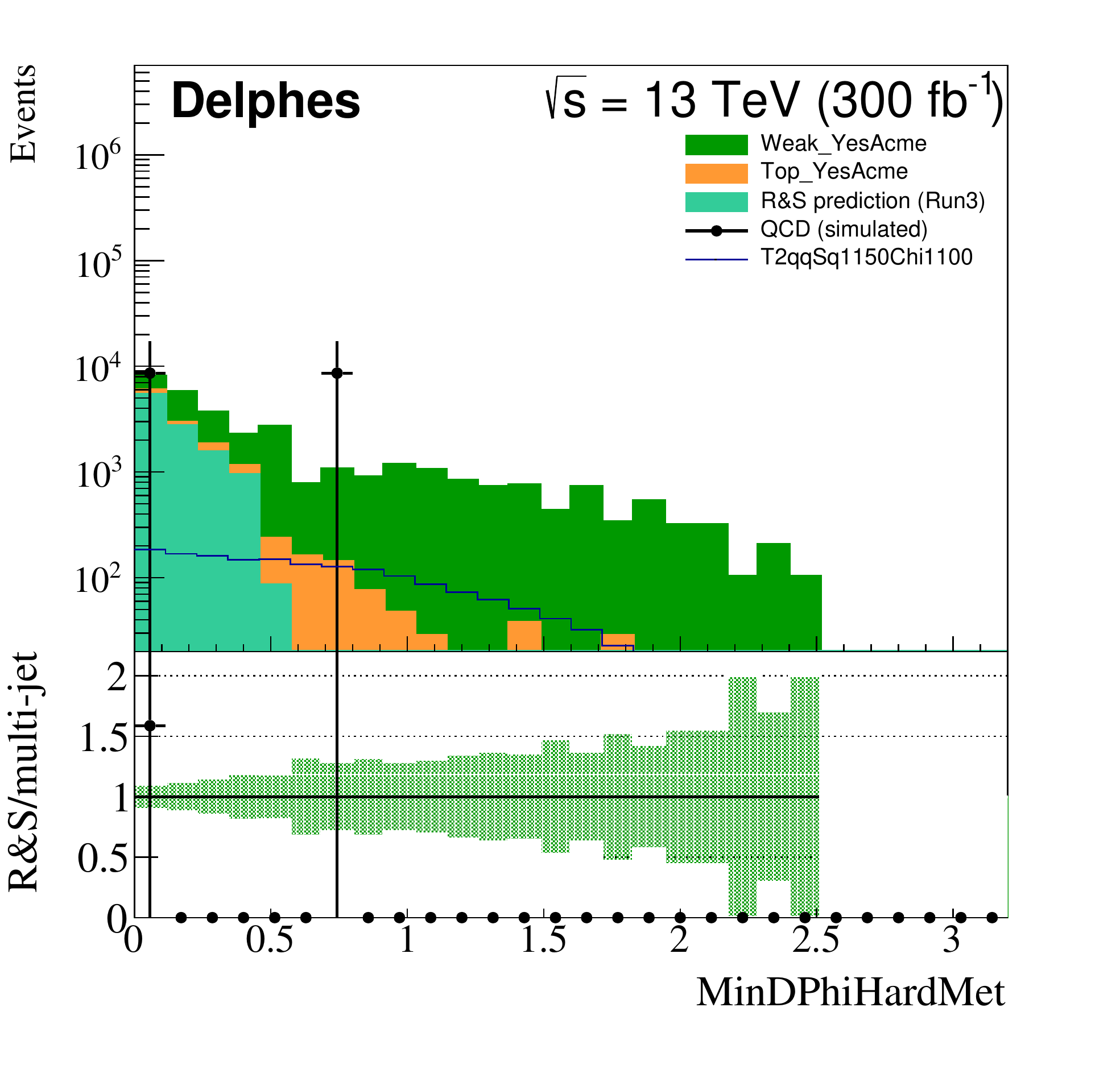}
}\\
\subfloat[]{
\includegraphics[width=0.45\linewidth]{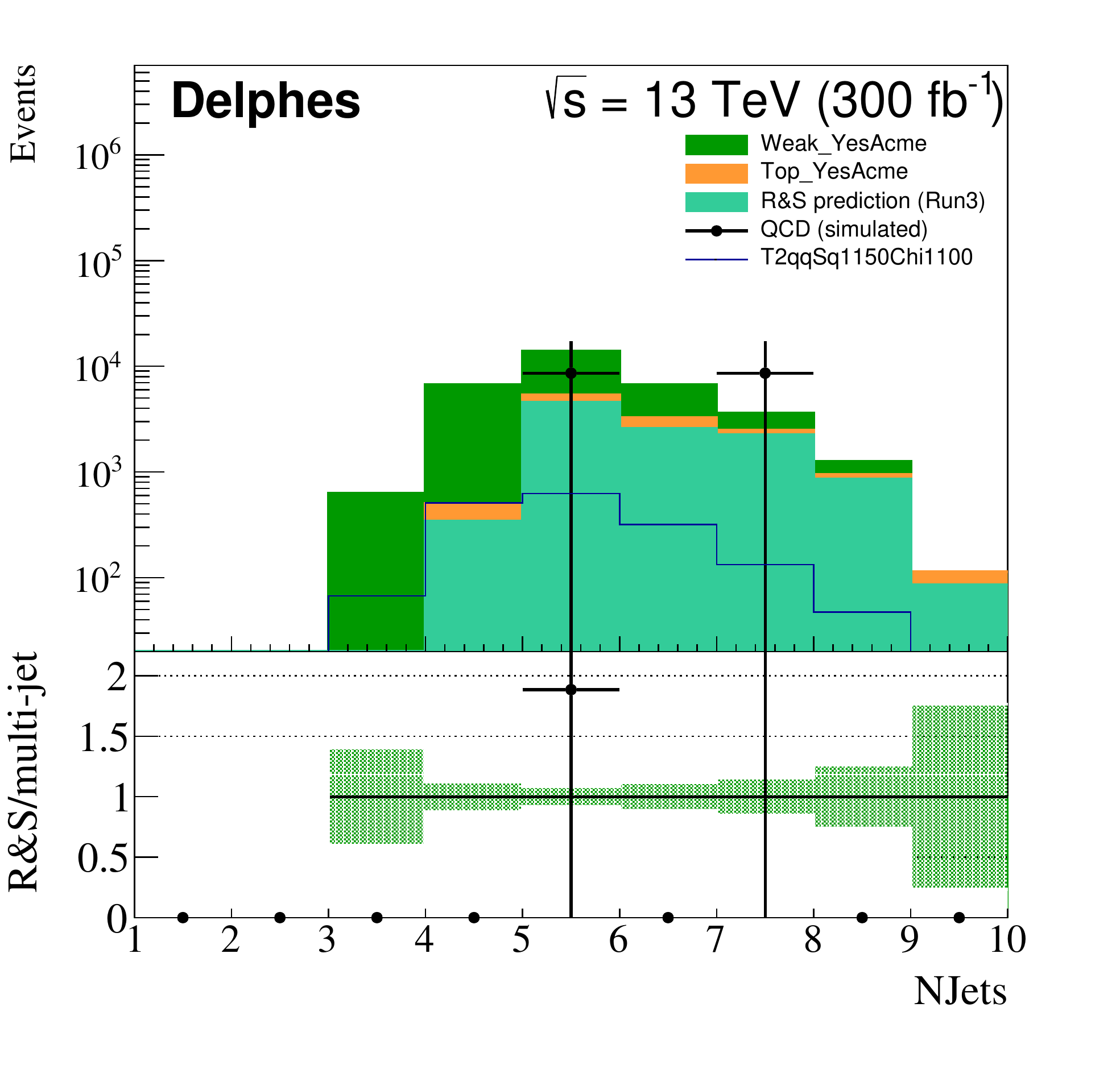}
}
\subfloat[]{
\includegraphics[width=0.45\linewidth]{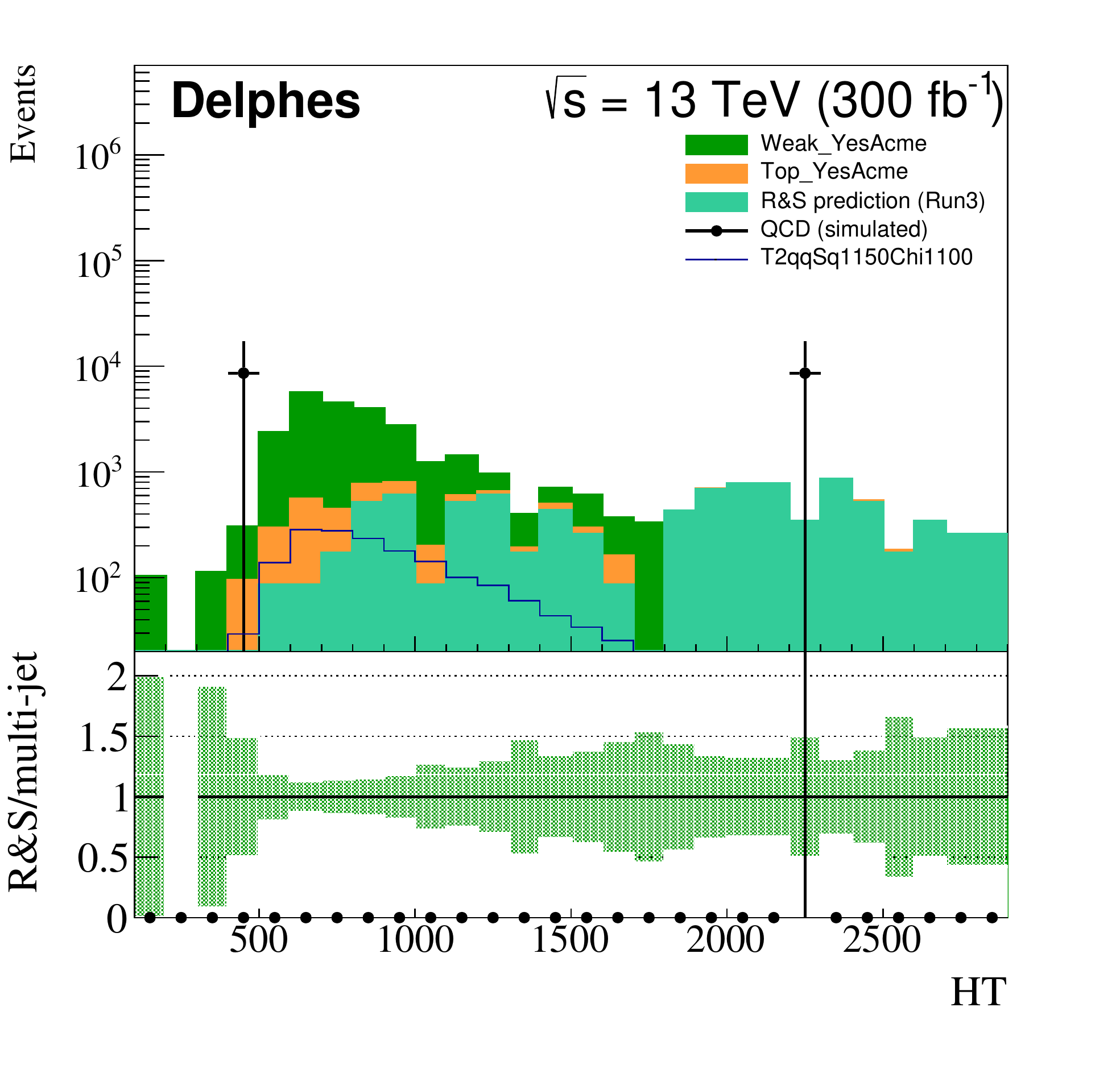}
}
\caption{Distributions of the analysis observables for signal and background events after the final event selection, namely, for events passing the pre-selection with a BDT score greater than 0.5.
The QCD background obtained directly from simulation is shown as black dots, while the QCD prediction obtained from the R\&S method, is shown in turquoise, and the lower panel indicates the ratio of the two. The QCD prediction is seen to give reasonable agreement with the simulated QCD.}
\label{fig:baselinebdt}
\end{figure}

The distributions of events passing the baseline selection are dominated by SM background, and it is clear that further purification is necessary.
The distribution of the classifier BDT is also shown for events passing the baseline selection in Fig. \ref{fig:baselinebdt}, along with other kinematic distributions passing a loosened BDT selection of BDT$>0.5$.  An advantage of the Rebalance and Smear technique can be seen from the fact that a statistically precise modelling of the QCD background is established in the signal region, even though only a handful of seed events exist to fill out this region. 

\clearpage

To select a final signal region, a scan is performed on the BDT threshold, and a tight cut of BDT$>$0.67 is found to yield a region with around 30\% signal purity, with still over 100 signal events surviving. Table \ref{tab:summary} provides a summary of the yields of the relevant processes, and estimate of the significance of the signal region. 


\begin{table}
\centering
\begin{tabular}{ l | c  c }
\hline
Process & SR (BDT$>$0.67) & $\sigma$(SR)   \\
\hline
QCD multi-jet &  81 $\pm$ 20 & \\
Non-QCD multijet &  279 $\pm$ 17 & \\
\hline
Total background &  360 $\pm$ 26 & \\
\hline
T2qq (1150,1100) &  125 $\pm$ 32 & 2.67 \\
\hline
\end{tabular}
\caption{Event counts and estimated uncertainties of the relevant processes after the signal region (SR) selection. The entry in the last column is a simplified significance calculated from the quoted yields and assumed uncertainties.}
\label{tab:summary}
\end{table}

The uncertainty in the non-QCD estimate is taken to be around 5\%, consistent with the lost-lepton and invisible $Z+\text{\,jets}$ background estimates made for signal regions with comparable background counts in Reference~\cite{Sirunyan:2019ctn}.
The signal uncertainty, corresponding to 25\%, is taken by summing in quadrature the sources listed in Reference~\cite{Sirunyan:2019ctn} for the scale, initial-state radiation, jet energy scale, jet energy resolution, pile-up modelling, trigger efficiencies, and $H_{T}$ and $H_{T}^{\text{miss}}$ modelling.
An uncertainty of 25\% is assigned to the R\&S QCD estimate to cover any possible non-closure, as well as typical uncertainties related to the jet response model.
A discussion of possible additional systematic uncertainties is given in the following. The significance $\sigma$ is computed in a simplified way as $\sigma=\frac{s}{\sqrt{s+b+(\delta s)^{2}+(\delta b)^{2}}}$.

%% file: uncertainties.tex
\section{Systematic uncertainty evaluation}
\label{sec:systematics}

Three main sources of systematic uncertainty have been identified, as well as a number of minor sources.
The main sources, as well as the methods considered for evaluating and assigning systematics are given in the following. 

\begin{enumerate}
\item Statistical:
The number of events in the seed sample is very large compared to the weighted count in the signal region. 
However, there is a probability that a single seed event enters the signal region more than once, and so the statistical uncertainty may not be that of a single Poisson distribution.
A boot strapping method is employed whereby the prediction is performed multiple times, each time based on a randomly selected subset of the seed sample.
In practice, as few as ten random subsets can be used and result in an ensemble of 10 estimates who's mean and RMS are taken as the central value and statistical uncertainty. 
\item Jet response:
The jet responses, used both in the rebalancing and smearing steps, are typically derived from simulation, and modified to match the data, with an associated uncertainty.
It is incumbent upon the particular experiment to characterize the uncertainty in the used jet response model.
Ideally, this uncertainty accounts for potential mis-modelling of the Gaussian core of the jet response, as well as the tails.
In practice, one or two variants of the jet response function, corresponding to a one-$\sigma$ widening of the Gaussian width or tail fraction of the jet response, are used to independently determine the QCD multi-jet prediction, and the variation in the predicted value is taken as a systematic. 
\item Non-closure:
A closure test based on simulation, such as that shown in the ratio panels of Figs. \ref{fig:baseline} and \ref{fig:baselinebdt}, can reveal any discrepancies that may be the result of assumptions of the method.
These assumptions include, but are not limited to, the factorisability of the prior into $\slashed{p}_{\text{T}}$ PDF and a $\Delta\phi(\slashed{\boldsymbol{p}}_{\text{T}}^{\text{hard}}, \boldsymbol{p}_{\text{T}}^{j1})$ PDF, or any rare non-convergent behaviour of the rebalancing fit. 
\end{enumerate}

%% file: summary.tex
\section{Summary}
\label{sec:conclusion}

The Rebalance and Smear method for estimating the fake-\met{} background for search regions with moderate or large \met{} has been presented in detail. The method, originally developed within the CMS experiment and deployed in both CMS and ATLAS, is found to be suitable for predicting the multi-jet backgrounds in a range of final states, including final states defined by cuts on or shapes of multivariate classifiers that correlate many event-level observables. We find that such classifiers may be needed to maximize the BSM programme of LHC searches, e.g., for exploring Higgsino dark matter scenarios, and so the method's utility may only be expected to increase.  We present a stand-alone methodology and tool set, consisting of software that interfaces with public tools such as \textsc{Delphes3}. Provided code computes and maximizes the posterior density, and performs both the rebalancing and smearing steps. Because of its simplicity and modularity, the code is adaptable to a specific experiment's or analysis' needs. The likelihood has also been developed to allow for a mixture of objects with large and small momentum resolution, i.e.\ jets, leptons, and photons. R\&S is a generative model, and its statistical power has been showcased, illustrating how the statistical precision of the seed sample can be boosted to an arbitrary degree through repeated smearing steps. Prescriptions for assigning systematic uncertainties have been suggested, including those necessary to cover errors associated with the repeated smearing.

%% file: appendix_code.tex
\section{Framework/code, simulation}
\label{app:code}

A framework for the implementation of Rebalance and Smear is maintained in \cite{RnSCode}. 
The core Rebalance and Smear functions are encoded in the header file 
\begin{center}
\texttt{BayesQcd/src/BayesRandS.h}.
\end{center}
 The library contains a number of global observables used to track the collections of original, rebalanced, and smeared jets, as well as parameters involved in the fitting procedure.
Noteworthy functions contained therein, and their main utility, are:
\begin{itemize}
 \item \textit{GleanTemplatesFromFile}: gather the $p_{T}$ and $\eta$ slices of the jet response PDF (likelihood), as well as the hard $E_{T}^{\text{miss}}$ prior into one object. 
  \item \textit{findJetToPin}: identifies a starting point for the rebalancing fit parameters. Pairs of jets are rescaled so as to result in a small $E_{T}^{\text{miss}}$. 
 \item \textit{fcn}: the rebalancing posterior density function determines the value of the posterior PDF for a particular arrangement of jets. 
 \item \textit{RebalanceJets}: returns a vector of rebalanced jets after running the posterior maximization.
 \item \textit{smearJets}:  returns a vector of rebalanced and smeared jets by applying a rescaling to the rebalanced jets derived via a random sampling of the jet response PDF for each jet. 
 \end{itemize}
These functions are roughly called in order within the skimming script 
\begin{center}
\texttt{tools/skimDataRebalanceAndSmear.py},
\end{center}
which provides a working example of running Rebalance and Smear over \textsc{Delphes3} simulated input data.
A \textsc{ROOT} TTree \textit{littletree} is constructed with a number of branches intended for use in the physics analysis.
Within the event loop, the quantities corresponding to the branches are computed from the input \textsc{Delphes3} collections, as well as from the rebalanced and smeared collections.
Both the original and rebalanced and smeared collections are saved interleaved within the same tree, and a binary branch \textit{IsRandS} identifies if a particular entry corresponds to an original or rebalanced and smeared event.
A simple auxiliary TTree \textit{tcounter} is defined with no branches and is filled once per event before any cuts, which serves to keep track of the total number of events analyzed. 

A histogram drawing script,
\begin{center}
\texttt{tools/DrawAnalyze.py},
\end{center}
is included that runs over the above produced skim and draw properly weighted histograms of desired quantities, from which figures as well as information for limit setting can be derived. 

A key input to the above analysis chain is the file used in the skimmer script as \textit{ftemplate}, which organizes the PDFs used for the likelihood and prior calculations.
This file must be built from scratch using properties of the jets and the detector of the project of interest, and is created by running the script
\begin{center}
\texttt{tools/LlhdPriorHistmaker.py}.
\end{center}
This script must be run over a high-statistics sample of simulated QCD multi-jet events in order to arrive at an appropriate set of prior and likelihood PDFs.
The output of this code is a set of histograms, which must be further processed using the script
\begin{center}
\texttt{tools/articulateSplines.py},
\end{center}
which performs a smoothing of the histograms into differentiable functions and endows them with an appropriate naming scheme. 

In total, $2.6\cdot 10^{6}$ QCD multi-jet, $3\cdot10^7$ $W+\text{\, jets}$ and $Z+\text{\, jets}$ boson, $2\cdot10^6$ $t\bar{t}$, simulated background MC have been generated using LO \textsc{Pythia8}, using the production keys \texttt{HardQCD:all = on}, \texttt{WeakBosonAndParton:all = on}, and \texttt{Top:all = on}, respectively.
In addition, $6\cdot 10^5$ signal events have been simulated using the same software and configuration.